\documentclass[aps,prd,amsfonts,preprint, eqsecnum,nofootinbib]
{revtex4} 
\usepackage{graphics,epsfig}
\begin{document}
\def\bullet{{(1)}}
\def\kkdl{d_L^\bullet} 
\def\kkdr{d_R^\bullet} 
\def\kkul{u_L^\bullet} 
\def\kkur{u_R^\bullet} 
\def\kkw{W^\bullet} 
\def\kkz{Z^\bullet} 
\def\kknu{\nu^\bullet} 
\def\kkl{l^\bullet} 
\def\kkb{B^\bullet} 
\def\kkdlbar{\bar{d}_L^\bullet} 
\def\kkdrbar{\bar{d}_R^\bullet} 
\def\kkulbar{\bar{u}_L^\bullet} 
\def\kkurbar{\bar{u}_R^\bullet} 
\preprint{WM-03-111}
%
\title{\vspace*{0.5in} Universal Extra Dimensions and Kaluza Klein 
Bound States \vskip 0.1in}
\author{Christopher D. Carone}\email[]{carone@physics.wm.edu}
\author{Justin M. Conroy}\email[]{jconroy@camelot.physics.wm.edu}
\author{Marc Sher}\email[]{sher@physics.wm.edu}
\author{Ismail Turan}\email[]{ituran@newton.physics.metu.edu.tr} 
\affiliation{Particle Theory Group, Department of Physics,
College of William and Mary, Williamsburg, VA 23187-8795}
\date{December 2003}
\begin{abstract}
We study the bound states of the Kaluza-Klein (KK) excitations of
quarks in certain models of Universal Extra Dimensions.  Such bound
states may be detected at future lepton colliders in the cross section
for the pair production of KK-quarks near threshold.  For typical
values of model parameters, we find that ``KK-quarkonia" have widths
in the 10 - 100 MeV range, and production cross sections of order a
few picobarns for the lightest resonances. Two body decays of the
constituent KK-quarks lead to distinctive experimental signatures. We
point out that such KK resonances may be discovered before any of the
higher KK modes.
\end{abstract}
\maketitle

\section{Introduction}\label{sec:intro}

The possibility of large extra dimensions has met considerable
scrutiny in recent years.  Sub-millimeter sized extra dimensions, in
which only gravity can propagate in the bulk, allows for a
reinterpretation of the hierarchy problem~\cite{Anton90}.  TeV-scale 
extra dimensions allow gauge and matter fields to propagate in the 
bulk as well, and have the virtue of allowing for an accelerated gauge
unification~\cite{Dienes}.  These and related scenarios are
well-motivated by string theory, where the existence of extra spatial
dimensions is necessary for the consistency of the theory.

The notion that the propagation of gauge and matter fields in the bulk
implies compactification radii of order a TeV$^{-1}$ follows from
consideration of precision electroweak constraints~\cite{ewc}.  
In the first types of models studied, at least
one Higgs field was assumed to be confined to an orbifold fixed point.
The vacuum expectation value (vev) of such a field necessarily results
in mixing between the $Z$ boson and its Kaluza-Klein (KK) excitations.
One's intuition from models with extra $Z'$ bosons and $Z$-$Z'$ mixing
suggests that the bounds on the first KK excitation will be of order a
TeV, with some reduction if the vev of the Higgs responsible for this
mixing is particularly small.

Universal extra dimensions (UED) were proposed as a way of
avoiding such tree-level contributions to precision electroweak
observables altogether~\cite{UED}.  In UED, all fields propagate in the bulk.
Conservation of KK number prevents mixing between KK and zero-mode
electroweak gauge bosons, so that the bounds described earlier are
avoided.  In the case of one extra dimension compactified on a $Z_2$
orbifold, a residual $Z_2$ symmetry of the effective four-dimensional
(4D) Lagrangian allows interactions only between even numbers of the
odd numbered KK modes.  This renders the lightest KK particle (LKP)
exactly stable.  Typical bounds on the scale of compactification,
$1/R$, are weakened to the collider bounds for the pair production of
KK states, or approximately $300$~GeV~\cite{bpsw}.  The possibility 
that the LKP is a dark matter candidate has also been 
investigated~\cite{Majumdar}.

In the absence of radiative corrections and electroweak symmetry
breaking, all KK modes at a given level would be exactly degenerate,
with masses given by $n/R$, where $n$ is a non-negative integer.
Electroweak symmetry breaking introduces small corrections to this
spectrum, with perhaps the exception of the KK excitations of the top
quark, since $m_{\rm top}$ is not necessarily much smaller than $1/R$.
A more sizable effect results from loop corrections to the KK mass
spectrum, which can be divided into two types~\cite{Cheng}.  There are
finite corrections, resulting from the propagation of bulk fields around
the compact dimension, which are insensitive to momentum scales above $1/R$.
There are also logarithmically divergent contributions that are localized at
the orbifold fixed points.  These renormalize the possible 5D 
Lorentz-violating interactions that exist at the fixed points and
alter the KK mass spectrum.  If we think of these interactions as 
counterterms, a renormalization condition must be chosen to fix their
finite parts.  Corrections to KK masses are thus determined by $1/R$, the
ultraviolet cutoff of the theory $\Lambda$, and the renormalization
condition that determines the finite parts of the fixed-point-localized
counterterms.  Although in the most general case, these finite parts
are undetermined (and the scenario is devoid of predictivity) one can
adopt a minimal assumption that they vanish at the cutoff 
$\Lambda$.  This boundary condition is no worse than, for example, the 
assumption of universal soft masses at the unification scale in the 
minimal supersymmetric standard model.  We will adopt this assumption for 
the present purpose, and will show later that our results do not strictly 
depend on it.

A consequence of an otherwise degenerate mass spectrum corrected by
loop effects is the possibility that some approximate degeneracies may
remain.  In particular, we note in the present work that the mass
difference between the Kaluza-Klein excitations of the quarks (which
we will refer to as KK-quarks, for brevity) and the LKP can be
relatively small, for reasonable choices of $R$ and $\Lambda$.  The
implication that we explore is the possible formation of KK-quark
bound states, and we investigate whether they may be discerned at
future electron-positron and muon colliders.  In the case of heavy
standard model quarks, it is well known that toponium bound states do
not form because the lifetime of the top quark is short compared to
the time scale associated with hadronization.  It is usually said that
this is a consequence of the heaviness of the top quark, but more
precisely, it is a consequence of the large top-bottom mass
difference.  In the UED scenario of interest, the lightest KK quarks
must decay to the (stable) LKP, and the phase space suppression leads
to a different conclusion, for a wide range of model parameters.  An
investigation of KK bound states is not merely a topic of academic
interest.  It is possible that the pair production of KK modes of the
first level may be accessible at colliders while that of the second
level may be kinematically out of reach.  Then the search for bound
states of the first KK modes via a threshold scan may be the quickest
approach to discovering additional interesting physics.

Our paper is organized as follows.  In the next section we give a
detailed review of UED, including the topic of radiative corrections
to the mass spectrum.  In Section~3, we discuss the criterion for the
formation of bound states, determine the model parameter space that is
consistent with this constraint, and compute the bound state spectrum.
In Section~4, we discuss the production and detection of
``KK-quarkonia" at electron-positron, and at muon colliders.  In
particular, we show that the bound state decays have a distinctive
signature that should allow easy discrimination from backgrounds.  In
the final section we summarize our conclusions.

\section{UED}\label{sec:ued}

In this section we review the derivation of the 4D Lagrangian assuming
one universal extra dimension.  We begin by considering the simplified
example of a U(1) gauge theory and then immediately generalize to
the full standard model gauge group. We focus on results that will be
used in the phenomenological analysis that follows.

Consider a 5D U(1) gauge theory with a fermion of unit charge $e_{5D}$
propagating in the bulk.  In 5D, the Clifford algebra is given by
\begin{equation}
\{ \Gamma^M \, , \, \Gamma^N \} = 2 g^{MN} \,\,\, ,
\end{equation}
where $\Gamma^\mu = \gamma^\mu$ and $\Gamma^5 = -i \gamma^5$.  Here
Roman indices run over all dimensions, while Greek indices run over
the familiar four. It follows that the 5D spinor fields $\Psi$ have
four components, like their four-dimensional counterparts.  However,
since $\gamma^5$ no longer purely anticommutes with the 5D Dirac
operator $i \Gamma^M \partial_M$, no chirality can be assigned to a
massless 5D spinor field.

We now compactify the theory on the orbifold $S^1/Z_2$.
Four-dimensional chirality is obtained by imposing the boundary
conditions $\Psi(x^\mu, y)= \Psi(x^\mu, y + 2 \pi R)$ and $\Psi(x^\mu,
y) = -\gamma^5 \Psi(x^\mu, -y)$.  This implies that there is a $Z_2$
even field $\Psi_+$ that is left-handed, and an odd field $\Psi_-$
that is right handed,
\begin{equation}
\Psi_+(x^M) = \sum_{n=0}^{\infty} \Psi^{(n)}_+(x^\mu) \cos(ny/R) \,\,
,\,\, \Psi_-(x^M) = \sum_{n=1}^{\infty} \Psi^{(n)}_-(x^\mu) \sin(ny/R)
\,\, .
\end{equation}
Since $\Psi_-$ has no $n=0$ component, only a left-handed zero-mode
remains, while each higher KK level is composed of a vector-like pair.
A 5D gauge field may be similarly decomposed
\begin{equation}
{\cal A}_\mu(x^M) = \sum_{n=0}^{\infty} {\cal A}^{(n)}_\mu(x^\nu)
\cos(ny/R) \,\, ,\,\, {\cal A}_5(x^M) = \sum_{n=1}^{\infty} {\cal
A}^{(n)}_5(x^\nu) \sin(ny/R) \,\, .
\end{equation}
This choice of $Z_2$ parities assures that an unwanted scalar
photon zero mode is also projected away by the orbifold boundary
conditions.

The 4D Lagrangian may be obtained by substituting these expansions into
the 5D action
\begin{equation}
S = \int d^5 x \, \left(\overline{\Psi}\, i \Gamma^M D_M \Psi - \frac{1}{4}
F_{MN} F^{MN}+ {\cal L}_{\rm gauge\,\,fixing}\right)\,,
\end{equation}
where $D^M = \partial^M - i e_{5D} {\cal A}^M$, and integrating over the
extra dimension $y$.  Terms quadratic in the $n^{th}$ mode
$\Psi^{(n)}$ or ${\cal A}^{(n)}_\mu$ in the 4D theory are then found
to be multiplied by a factor of $2 \pi R$ if $n=0$, or $\pi R$ if
$n=k>0$.  Thus, properly normalized kinetic terms are obtained only
after the rescalings
\begin{equation}
{\cal A}^{(0)}_\mu = \frac{1}{\sqrt{2 \pi R}} A^{(0)}_\mu
\,,\,\,\,\,\, {\cal A}^{(k)}_M =\frac{1}{\sqrt{\pi R}} A^{(k)}_M
\,,\,\,\,\,\, \Psi^{(0)}_+ = \frac{1}{\sqrt{2 \pi R}} \psi^{(0)}_+
\,,\,\,\,\,\, \Psi^{(k)}_\pm = \frac{1}{\sqrt{\pi R}} \psi^{(k)}_\pm
\,\, .
\end{equation}
Notice that the fields $\Psi$ and ${\cal A}_\mu$ have mass dimensions
$2$ and $3/2$, respectively, while the rescaled fields $\psi$ and
$A_\mu$ have their usual 4D mass dimensions.  Taking these rescalings
into account, and that derivatives with respect to $y$ become factors
of $n/R$ in the $4D$ theory, one may easily find the tree-level masses
\begin{equation}
m_{\psi^{(n)}} = m_{A^{(n)}} = n/R \,\,\, .
\end{equation}

The gauge field fermion interactions for the $Z_2$-even fermion fields
follow from the 5D term
\begin{equation}
e_{5D} \, \overline{\Psi}_+ {\cal A}_\mu \gamma^\mu \Psi_+\,\,\,.
\end{equation}
Integrating over $y$, one finds
\begin{equation}
\!\!\!\!\!e_{5D} \left(\,2 \pi R\overline{\Psi}^{(0)}_+ \not \!\! 
{\cal A}^{(0)} \Psi^{(0)}_+ +\!\sum_{n>0} \pi R 
\overline{\Psi}^{(0)}_+ \not \!\!
{\cal A}^{(n)} \Psi^{(n)}_+ +\!\!\sum_{m>0,n,r} (\delta_{m,|n-r|} +
\delta_{m,n+r}) \, \frac{\pi R}{2} \overline{\Psi}^{(m)}_+ \not \!\!
{\cal A}^{(n)} \Psi^{(r)}_+\,\right)\,.
\end{equation}
Of relevance to our investigation of KK quark decays later are the
gauge interactions involving $n=0$ and $n=1$ modes.  With the field
rescalings described above, and including the $Z_2$ odd fermion field
one finds 
\begin{equation}
{\cal L} = e \left( \overline{\psi}^{(0)}_+ \not \!\! A^{(0)}
\psi^{(0)}_+ + \overline{\psi}^{(1)}_+ \not \!\! A^{(0)} \psi^{(1)}_+
+ \overline{\psi}^{(1)}_- \not \!\! A^{(0)} \psi^{(1)}_- +
[\overline{\psi}^{(0)}_+ \not \!\! A^{(1)} \psi^{(1)}_+ + h.c.]
\right)\,,
\end{equation}
where the 4D gauge coupling $e = e_{5D}/\sqrt{2 \pi R}$.  Note that
the 5D gauge coupling $e_{5D}$ has mass dimension $-1/2$, while $e$ is
dimensionless, as we expect. This expression may be written more
compactly be embedding the left- and right-handed modes $\psi_+$ and
$\psi_-$ into Dirac spinors $\psi$  
\begin{equation}
{\cal L}_{\rm LH} = e \left(\overline{\psi}^{(0)} \not \!\! A^{(0)}
P_L \psi^{(0)} + \overline{\psi}^{(1)} \not \!\! A^{(0)} \psi^{(1)}
+ [\overline{\psi}^{(0)} \not \!\! A^{(1)} P_L 
\psi^{(1)} + h.c.] \right)\,.
\label{eq:lhlag}
\end{equation}
Here $P_L=(1-\gamma^5)/2$, and the right-handed component of the
zero-mode Dirac spinor is arbitrary.  A similar expression for a
fermion with a right-handed zero-mode can be obtained from 
Eq.~(\ref{eq:lhlag}) by replacing $P_L$ by $P_R$.  If radiative 
corrections render $m_{\psi^{(1)}} > m_{A^{(1)}} + m_{\psi^{(0)}}$
than the last term can lead to KK fermion decay. 

The field rescalings and the KK mode numbers in Eq. (\ref{eq:lhlag})
are all independent of the chosen gauge group.  We therefore may
immediately generalize to the standard model.  The interaction terms
relevant to the KK-quark decays that we consider later are as follows:
\begin{eqnarray}
{\cal L} &=& \frac{2}{3} e \left[
\frac{\sin(\theta_W+\theta_{1})-\frac{1}{2}
\sin(\theta_W-\theta_{1})}{\sin2\theta_W}
\overline{u}^{(0)} \not\!\! A^{(1)} P_L u^{(1)}_L +
\frac{\cos\theta_{1}}{\cos\theta_W} \overline{u}^{(0)} \not\!\!
A^{(1)} P_R u^{(1)}_R \right] \nonumber \\ &+&\frac{1}{3} e \left[
\frac{\cos(\theta_W-\theta_{1})+
2\cos(\theta_W+\theta_{1})}{\sin2\theta_W} \overline{u}^{(0)} \not\!\!
Z^{(1)} P_L u^{(1)}_L -2 \frac{\sin\theta_{1}}{\cos\theta_W}
\overline{u}^{(0)} \not\!\! Z^{(1)} P_R u^{(1)}_R \right] \nonumber \\
&-&\frac{1}{3} e \left[ \frac{\sin(\theta_W+\theta_{1})- 2
\sin(\theta_W-\theta_{1})}{\sin2\theta_W} \overline{d}^{(0)} \not\!\!
A^{(1)} P_L d^{(1)}_L + \frac{\cos\theta_{1}}{\cos\theta_W}
\overline{d}^{(0)} \not\!\! A^{(1)} P_R d^{(1)}_R \right] \nonumber \\
&-& \frac{1}{3} e \left[ \frac{2\cos(\theta_W-\theta_{1})+
\cos(\theta_W+\theta_{1})}{\sin2\theta_W} \overline{d}^{(0)} \not\!\!
Z^{(1)} P_L d^{(1)}_L - \frac{\sin\theta_{1}}{\cos\theta_W}
\overline{d}^{(0)} \not\!\! Z^{(1)} P_R d^{(1)}_R \right] \nonumber \\
&+& \frac{1}{\sqrt{2}} e \left[
\frac{1}{\sin\theta_W}\overline{u}^{(0)} \not\!\! W^{(1)} P_L
d^{(1)}_L \right]+ h.c.
\label{eq:theints}
\end{eqnarray}
Note that the $n=1$ fields above are complete Dirac spinors (with both
left- and right-handed components), and the subscript indicates only
the chirality of the associated zero mode. In addition, $\theta_W$ is the
zero-mode weak mixing (Weinberg) angle, while $\theta_{1}$ is the
corresponding angle for the $n=1$ modes.  In the absence of radiative
corrections, the electroweak symmetry conserving contributions to the
$B^{(1)}$-$W^{(1)}$ mass squared matrix are precisely diagonal (and
equal to $1/R^2$), so that we expect $\theta_W=\theta_{1}$.  In that
limit, the photon and Z couplings in Eq.~(\ref{eq:theints}) have the
same values as their couplings to either left- or right-handed up or
down quarks.  However, radiative corrections lead to much smaller
values of $\theta_{1}$.  For example, for $\Lambda R=20$ and
$R^{-1}=500$~GeV, $\sin^2\theta_1 \approx 10^{-2}$~\cite{Cheng}.  In
the following section, we omit the dependence on $\theta_1$ to
streamline our analytical expressions.  The full dependence on
$\theta_1$ has been taken into account in all our numerical results,
and complete analytical expressions are provided in the Appendix.

\begin{figure}[htb] 
\vskip -3.3cm \centerline{ \epsfxsize 4.5in
{\epsfbox{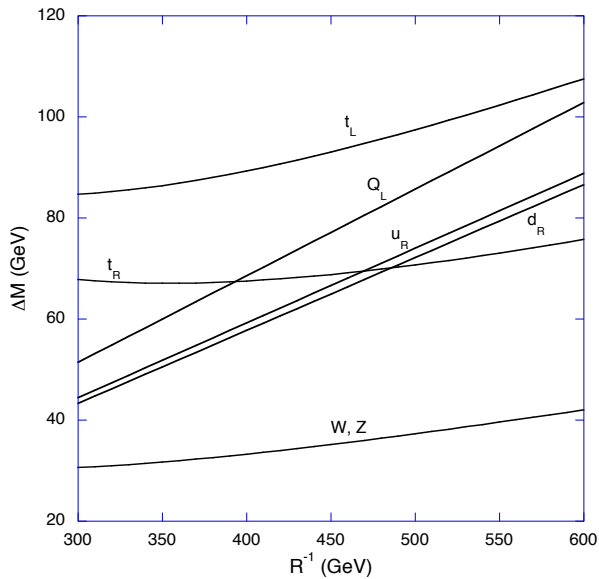}}}
    \vskip -3.6cm \caption{The mass splitting between KK-quarks
and the LKP, $\gamma^{(1)}$, as well as the splitting between the weak
KK-gauge bosons and the LKP, as a function of $1/R$ for $\Lambda
R=20$. Here, $Q_L$ stands for all isodoublet KK-quarks except top,
$u_{R}$ for up and charm isosinglet KK-quarks, and $d_{R}$ for down,
strange and bottom isosinglet
KK-quarks}\label{masssplittings}\end{figure}

Radiative corrections to the KK-gauge boson and KK-quark masses allow
for two-body decays via the interactions in Eq. (2.11). Over the range
of $\Lambda R$ and $R^{-1}$ that we consider, the LKP is the first KK
excitation of the photon, $\gamma^{(1)}$~\cite{Cheng}. The radiative
corrections to the KK-quark and the KK-gauge boson masses were
calculated by Cheng, Matchev and Schmaltz \cite{Cheng}. Adopting their
assumption that the finite parts of counter terms vanish at the cutoff
scale $\Lambda$, we plot the mass splitting between the KK-quarks and
the LKP, as well as the splitting between the weak KK-gauge bosons and
the LKP, as a function of $1/R$, in Fig.~\ref{masssplittings}, setting
the value of $\Lambda R = 20$.  Complete expressions for the radiative
corrections that are taken into account in this figure can be found in
Ref.~\cite{Cheng}.  As a consequence of the smallness of the $n=1$
mixing angle $\theta_1$, the LKP, $\gamma^{(1)}$, is almost entirely a
KK-$B$ boson, while the KK-W and KK-Z are virtually degenerate in
mass.  As we will see in the next section, the values of $\Delta M$ in
Fig.~\ref{masssplittings} are small enough to lead to KK-quark bound
state formation.

\section{Bound States}\label{sec:bs}
From Fig.~\ref{masssplittings}, one finds that radiative corrections
to the KK masses in UED are typically in the 10-100 GeV range. We will
show that this is numerically small enough to allow for the formation
of bound states of KK quarks.  The smaller phase space available for
KK-quark decay renders the bound states narrower than the spacing
between adjacent KK-quarkonia levels, at least for the first few
levels. In this section, the decay widths and branching ratios of the
KK-quarks are calculated and discussed, as well as the mass splittings
of the different KK-quarkonia energy levels and the production cross
sections at lepton colliders.

\subsection{Decay widths and branching ratios} 

With the Lagrangian and mass splittings given in the previous section,
it is straightforward to determine the decay widths and branching
ratios of the KK-quarks.  We will begin by considering the weak
isosinglet KK-quarks (except the KK-top), then the weak isodoublet
KK-quarks, and finally the unusual case of the isosinglet KK-top
quark. While the partial decay widths of KK-quarkonia through
annihilation are typically tens of keV, we will see that the decay
widths of the KK-quarks (except the KK-top) are typically close to a
hundred MeV. Thus, the decay width of a KK-quarkonium state will be
twice the decay width of the KK-quark.

\begin{figure}[htb] 
\centerline{ \epsfxsize 4.5in 
{\epsfbox{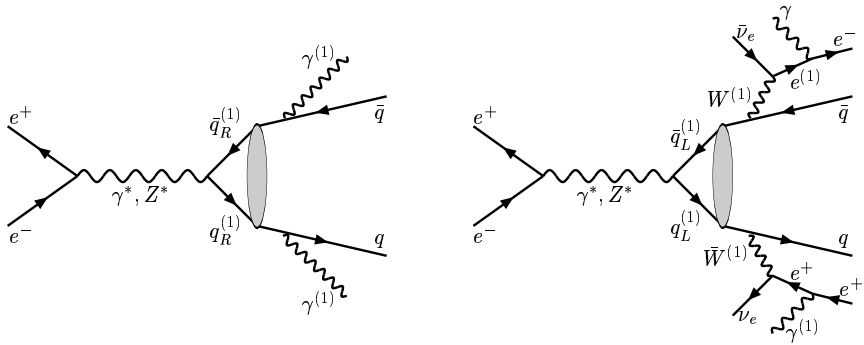}}}
\caption{The production and decay chains of $q_R^{(1)}$ 
and $q_L^{(1)}$ pairs.  Note that all of the decays in the $q_L^{(1)}$
decay chain are two-body, leading to monochromatic quarks and leptons.}
\label{newfig}\end{figure}

\subsubsection{Isosinglet KK-quarks}
 
\begin{figure}[h]
\vskip -3.3cm \centerline{ \epsfxsize 4.5in
    {\epsfbox{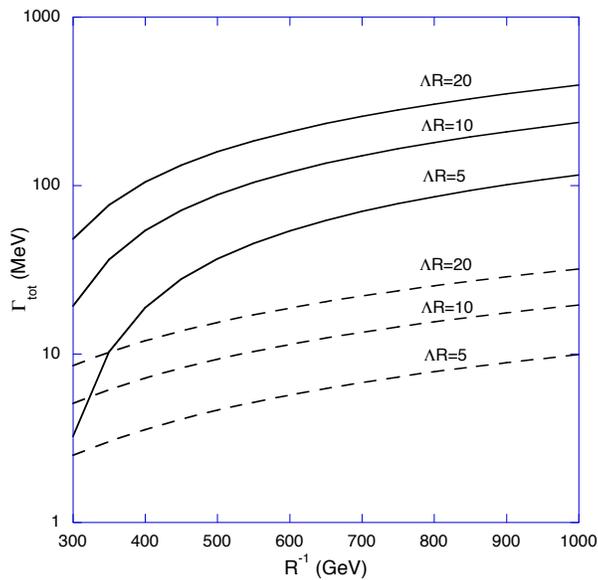}}}
 \vskip -3.6cm \caption{The total decay width of $n=1$ isosinglet and
isodoublet down KK-quarks as a function of $R^{-1}$ for fixed $\Lambda
R=5,\,10,\,20$.  The solid lines represent the total decay widths of
isodoublet down KK-quarks for each corresponding $\Lambda R$ value,
respectively. The dashed lines are for the isosinglet case.  The
isodoublet up KK-quark total decay width is equal to that of the down
and the isosinglet up KK-quark's width is four times larger than that
of the isosinglet down.}\label{leftwidths}
\end{figure}

Isosinglet KK-quarks cannot decay into KK-W bosons and their decay
into KK-Z bosons is suppressed by a factor of $\sin^{2}\theta_1$. In
addition, their decay into KK-Higgs bosons is suppressed by small
Yukawa couplings.  As a result, the dominant decay mode is $q^{(1)}\to
q^{(0)}\,\gamma^{(1)}$, as shown in Fig.~\ref{newfig}.  Since the LKP
is stable, the decay signature will be a monochromatic quark and
missing energy.  The one exception is the isosinglet KK-top quark,
which cannot decay into a top quark and the LKP; we will discuss that
case shortly.  Including the small coupling to the KK-Z boson, we find
that the branching ratio into a quark and a KK-$\gamma$ is over 98
percent (consistent with the results in \cite{Cheng66}). Neglecting
the mass of the light quark and $\sin^2\theta_1$, we find the decay
width
\begin{equation} \Gamma(\kkdr\rightarrow d_{R} \, \gamma^{(1)})
={e^{2}m_{\kkdr}\over
288\pi\cos^{2}\theta_{W}}\left( 1 - {m_{\gamma^{(1)}}^{2}\over
m_{\kkdr}^{2}}\right)^{2}\left( 2 + {m_{\kkdr}^{2}\over
m_{\gamma^{(1)}}^{2}}\right)\,. 
\end{equation} 
An exact expression is given in the Appendix. The decay width for the
$\kkur$ is larger by a factor of four.  Given values for $1/R$ and
$\Lambda R$, this width is completely determined.  The results are
shown in Fig.~\ref{leftwidths}. We see that the widths are typically
within a factor of two of $10$ MeV. As noted above, the decay
signature is a monochromatic quark and missing energy; for $1/R=500$
GeV and $\Lambda R = 20$, the quark energy is $67$ GeV.
\subsubsection{Isodoublet KK-quarks} 

For the isodoublet KK-quarks, decay channels into KK-W and KK-Z bosons
are available.  Although there is less phase space into these than
into the KK-$\gamma$ boson, the couplings are substantially larger,
and the KK-W and KK-Z modes dominate.  The decay width into a KK-W is
given by 
\begin{equation} \Gamma(\kkdl\rightarrow u_{L}\,\kkw)={e^{2}m_{\kkdl}\over
64\pi\sin^{2}\theta_{W}}|V_{ij}|^2\left( 1 - {m_{\kkw}^{2}\over
m_{\kkdl}^{2}}\right)^{2}\left( 2 + {m_{\kkdl}^{2}\over
m_{\kkw}^{2}}\right)\,, 
\end{equation} 
where $V_{ij}$ is the relevant CKM element.  The identical decay
width, of course, applies to $\kkul\to d_{L}\,W^{(1)}$ decays.  The
decay width into a KK-Z is given by
\begin{equation} 
\Gamma(\kkdl\rightarrow d_{L}\, \kkz)={e^{2}m_{\kkdl}\over
128\pi\sin^{2}\theta_{W}}\left( 1 - {m_{\kkz}^{2}\over
m_{\kkdl}^{2}}\right)^{2}\left( 2 + {m_{\kkdl}^{2}\over
m_{\kkz}^{2}}\right)\,. 
\end{equation} 
We see that the branching ratio into the KK-W bosons are $2|V_{ij}|^2$
times that of the KK-Z bosons.  The branching ratio into KK-$\gamma$
bosons is negligible, always less than a few percent.  These decays
will give spectacular signatures.  The decay into a KK-W boson, as
shown in Fig.~\ref{newfig}, leads to the decay chains
$\kkdl\rightarrow u_{L}\,\kkw\rightarrow u_{L}\,l\, \kknu$ and
$\kkdl\rightarrow u_{L} \, \kkw\rightarrow u_{L}\,\kkl\, \nu
\rightarrow u_{L}\, l\, \nu \, \gamma^{(1)}$ leading to a
monochromatic quark, a lepton, and missing energy.  For example, for
$1/R=500$ GeV and $\Lambda R=20$, the quark energy will be $46$
GeV. Assuming measurement of the quark jet allows reconstruction of
the $W^{(1)}$ four-momentum, then the lepton energy can be completely
determined\footnote{Even if the $W^{(1)}$ four-momentum can't be
reconstructed, the spread in the lepton energy will be ${\cal
O}(10\%)$.}.  The decay into a KK-Z is even more spectacular, with the
chain $\kkdl\rightarrow d_{L}\, \kkz\rightarrow d_{L}\, l\, \kkl
\rightarrow d_{L}\, l\, l\, \gamma^{(1)}$, where $l$ is a charged
lepton.  Again, the initial quark jet energy is fixed, and the
sequential two-body decays should allow for easy reconstruction of the
event, and suppression of backgrounds.  Of course, in both the KK-W
and KK-Z cases, there will also be hadronic decays -- we have focused
on the leptonic because the signatures are much cleaner.  The
resulting total widths are plotted in Fig.~\ref{leftwidths}; and the
KK-W final state accounts for 2/3 of the widths (for the first two
generations).
 
For the first two generations, generation-conserving decays are favored,
since the CKM matrix is nearly the identity.  However, for the third
generation, a decay into a top quark is not kinematically allowed. For
the KK-top quark, this means that only the decay into a KK-W is
possible.  Due to CKM suppression, decays of the KK-bottom into KK-Z
and KK-$\gamma$ are favored and thus the decay width of the KK-bottom
is 1/3 of those shown in Fig.~\ref{leftwidths}.  For the isodoublet
KK-top, the mass is somewhat larger than the other KK-quarks, and thus
more phase space is available.  For most of parameter-space, we find
that the decay width of the KK-top (entirely through the KK-W chain)
is approximately 80 percent of the widths shown in the figure.

\subsubsection{isosinglet KK-top quarks} 

The isosinglet KK-top quark has a very different decay signature.
Flavor-conserving decays are kinematically forbidden, while
flavor-changing decays are not possible since there is no coupling to
the KK-W.  Hence, there are no two-body decays at tree-level.  The
isosinglet KK-top can decay into a KK-$\gamma$ and a virtual top
quark, which then decays into a bottom quark and a virtual $W$, which
then decays.  This four-body decay will be strongly phase-space
suppressed.  One expects four-body phase space to be roughly $10^{-4}$
of two-body phase space, and this leads to a decay width of the
isosinglet KK-top of approximately $1-10$ keV.  This is small enough
that the annihilation of the KK-quarkonia will dominate.  As we will
see in the next section, the annihilation width into electron pairs
will be $10-20$ keV, and when the other annihilation channels (other
leptons and quarks) are included, the total width will be well over a
hundred keV.  Thus, the dominant decay mechanism will be primarily
into fermion-antifermion pairs, and there will be no missing energy in
the decay.

\subsection{Production cross-sections} 
The cross-section for production of a vector resonance, 
$e^+e^-\rightarrow V\rightarrow X$ is given by \cite{Barger} 
\begin{equation} \sigma_V = {12\pi (s/M^2_V) \Gamma_{ee}\Gamma_X \over 
    (s-M^2_V)^2 + M_V^2 \Gamma_V^2}\,\,, 
\end{equation} 
 where $\Gamma_{ee},\Gamma_X$ and $\Gamma_V$ are the 
partial widths for $V\rightarrow e^+e^-$, for $V\rightarrow X$ and for the 
total width, respectively. Since we are interested in the total production 
cross-section, and since the partial width into $\Gamma_{ee}$ is much 
smaller than the total width, we can set $\Gamma_X=\Gamma_V$ 
(this will be valid for all cases except the isosinglet KK-top quarkonia).  
At the peak resonance, the production cross-section is then given by 
\begin{equation} \sigma_V^{\rm peak} = 
    {12\pi\over M^2_V}{\Gamma_{ee}\over \Gamma_V}\,\,. 
\end{equation}
We need the partial decay width of $V\rightarrow e^+e^-$.  
The decay width through a virtual photon is given by 
\begin{equation} \Gamma(V\rightarrow \gamma^* \rightarrow e^+e^-) \equiv 
    \Gamma_\gamma = {4\pi\alpha e_Q^2 \over 3 M^3_V}|F_V|^2\left[ 1 - 
    {16\alpha_s\over 3\pi}\right]\,, 
\end{equation} 
where $|F_V|^2$ is related to the wave function at the origin, and is given by 
$12 M_V |\Psi_V(0)|^2$.  The partial decay width including virtual 
Z exchange is related to this 
\begin{equation} \Gamma(V\rightarrow \gamma^*,Z^* \rightarrow e^+e^-) = 
    (M^2_V/e^2e_Q)^2(|G_V|^2+|G_A|^2)\Gamma_\gamma\,\,, 
\end{equation} 
where 
\begin{equation} G_V = {e^2e_Q\over M^2_V} + {8G_FM_Z^2\over\sqrt{2}}{g_V^e 
    g_V^Q \over M^2_V-M^2_Z+i\Gamma_Z M_Z}\,\,, 
\end{equation} and 
\begin{equation} G_A = {8G_FM_Z^2\over\sqrt{2}}{g_A^e g_V^Q \over 
    M^2_V-M^2_Z+i\Gamma_Z M_Z}\,\,. 
\end{equation} Here, $g_V$ and $g_A$ are the vector and 
axial vector couplings of the fermion to the Z, and  
$g_{V_{L}}^{Q}=g_{Q_L}\,,\; g_{V_{R}}^{Q}=(g_{Q_L}+g_{Q_R})/2$ with
$g_{Q}=T_{3}-e_{Q}\sin^2{\theta_W}$. Finally, we need the wave function 
at the origin.  At these high mass scales, one expects single gluon 
exchange to be fairly accurate, and in that approximation the wave function 
at the origin is given by 
\begin{equation}     
|\Psi(0)|^{2}={1\over\pi}\left({2 M_{Q}\alpha_{s}(M_{Q})\over 
3n}\right)^{3}\,\,,     
\end{equation} 
where $n$ is the principle quantum number. Putting these together, we find that 
\begin{equation}
\displaystyle
\Gamma_{ee}=\frac{e^4 \alpha_{s}^3(M_Q)}{27n^3
\pi^2}\,M_{V}\left(1-\frac{16\alpha_{s}}{3\pi}\right)
\left(e_{Q}^2+\frac{(g_{V}^Q)^2}{(1-\kappa_{Z})^
2\sin^4{2\theta_W}}\right)\,,\label{GamV} 
\end{equation}
where $\kappa_{Z}=m_{Z}^2/M_{V}^2$. 
\begin{table}[t]
\vskip -1.7cm
\caption{The partial decay width of $V\rightarrow e^+e^-$ for both 
isodoublet and isosinglet KK-quark bound states .} \label{Table}     
\begin{center}
\begin{tabular}{ccccc}
    \hline\hline
$M_{V}$(GeV) &$\;\Gamma_{ee}(\kkulbar\kkul)$(keV)
& $\;\Gamma_{ee}(\kkdlbar\kkdl)$(keV) &
    $\;\Gamma_{ee}(\kkurbar\kkur)$(keV) & 
$\;\Gamma_{ee}(\kkdrbar\kkdr)$(keV)\\ \hline 
    $600$       &14.58      &       6.73 &  9.74  & 3.64 \\
    $800$       &19.31      &       8.79 &  12.97 & 4.82 \\
    $1000$      &24.06      &       10.89&   16.21 & 6.01 \\
    $1200$      &28.82      &        13.00 &   19.45  & 7.20 \\
    $1400$      &33.59      &        15.13 &   22.68  & 8.39 \\
    $1600$      &38.36      &        17.25 &   25.92  & 9.59 \\
    $1800$      &43.14      &        19.39 &   29.16  & 10.78\\
    $2000$      &47.92      &        21.52 &   32.40  & 11.98\\
    \hline \hline
        \end{tabular}
        \end{center}
\end{table}
\begin{figure}[htb] 
\vskip -4.0cm
\centerline{ \epsfxsize 4.5in {\epsfbox{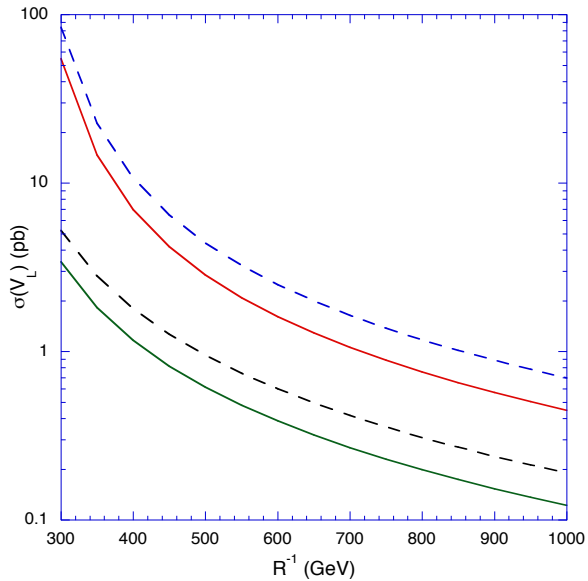}}}
    \vskip -3.6cm   
\caption{The resonance production cross section for isodoublet
KK-quarkonia states, except KK-toponium, as a function of $R^{-1}$
for $\lambda R=5,\,20$. The solid lines represent down type
KK-quarkonia states and the dashed ones represent up-type KK-quarkonia
states, and the upper (lower) lines correspond to $\lambda R=5 \, (20)$.}
\label{CrosSecL}\end{figure} 
\begin{figure}[htb]
\vskip -4.8cm
    \centerline{ \epsfxsize 4.5in {\epsfbox{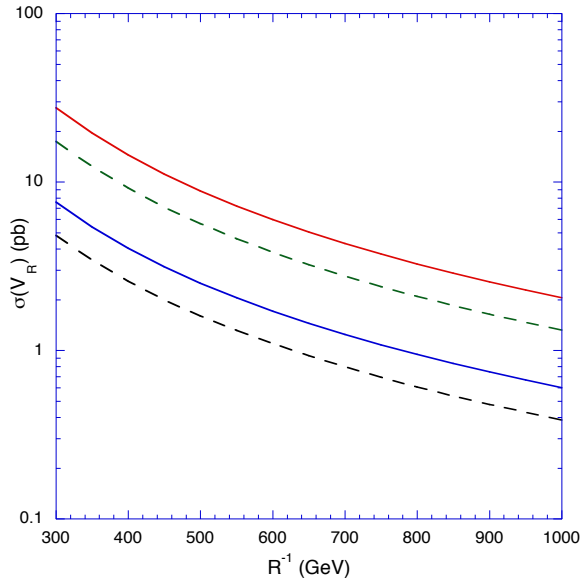}}}
 \vskip -3.6cm \caption{The same as Fig.~\ref{CrosSecL} but for
 isosinglet KK-quarkonia states. Here the upper (lower) lines
 correspond to $\lambda R=20 \, (5)$.} 
\label{CrosSecR}
\end{figure}
To get Eq. (\ref{GamV}) we assumed that $g_{V}^e$ is negligible and
$\Gamma_{Z}^2\kappa_{Z}\ll (1-\kappa_Z)^2 M_{V}^2$\,, even though we
have used the exact expressions for numerical calculations.  In Table
\ref{Table}, we have listed the decay width, $\Gamma_{{ee}}$ for a
range of KK-quarkonia masses.

We can now determine, using Eq. (3.5), the peak production cross
sections for isodoublet and isosinglet KK-quarkonia.  The results are
show in Figs.~\ref{CrosSecL} and \ref{CrosSecR}.  The cross sections
are substantial, between 1 and 100 picobarns.  

We now turn to the mass splittings between the different KK-quarkonia 
levels.

\subsection{Mass Splittings}
To observe KK-quark bound states, the mass splitting between adjacent
resonances must be larger than their typical decay widths.  The
KK-quark mass scale justifies a non-relativistic calculation of the
binding energies. We therefore solve the radial Schr\"odinger
equation,
\begin{equation}
-\frac{1}{2\mu}\frac{d^{2}u}{d
 r^{2}}+[V(r)+\frac{1}{2\mu}\frac{l(l+1)}{r^{2}}]\,u=\Delta E \, u\,, 
\label{eq:schr}
\end{equation}
for a suitable phenomenological potential $V(r)$. Here $u(r)=r R(r)$
and the complete wave function is $\psi(r,\theta,\phi)= R(r) Y_{l
m}(\theta,\phi)$. The wave function satisfies
\begin{eqnarray}
 &u(0)=0\,,\nonumber \\
 &u(r) \rightarrow 0\,,\; r\rightarrow \infty\,. 
\end{eqnarray}
Given a choice for $V(r)$, Eq.(\ref{eq:schr}) is
solved numerically to obtain the energy eigenvalues $\Delta E$;
the mass for each bound state is then given by 
\begin{equation}
M_{n}=2M_{KK}+\Delta E_{n}\,,
\end{equation}
where $\Delta E_{n}$ is the energy eigenvalue of the $n^{\mbox{th}}$ level 
and $M_{KK}$ is the mass of the KK-quark.  Normalizing the wave function by
\begin{equation}
\int_{0}^{\infty} |u|^2 dr=1 \,\, ,
\end{equation}
one may compute $R(0) = u'(0)$.

QCD motivates the following form for the potential:
\begin{equation}
V=-\frac{4}{3}\frac{\alpha_{s}}{r}+A r\,. \label{eq:V} 
\end{equation}
    
The first term is Coulomb-like and is generated by one-gluon exchange, 
while the second is linear and models confinement. For
$A \simeq 1$~GeV $\mathrm{fm}^{-1}$, this potential predicts energy
level splittings in good agreement with the data for the $\Upsilon$
and J/$\psi$ systems.  At the typical energies of KK-quarkonia
production, one would expect the Coulomb-like potential to dominate
resulting in nearly hydrogen-like energy level splittings.  However,
hydrogen-like wave functions become more spread out at higher energy
levels, suggesting a more significant contribution from the linear
term in these cases.  Note, however, that the level spacings
for a hydrogen-like spectrum decrease roughly as $\displaystyle \Delta
E_{n,n+1}\propto \frac{1}{n^{3}}$\,, where $n$ is the radial quantum
number.  Therefore, only the first few energy levels will have
splittings large enough to permit KK-quark bound states to be
distinguished.

\begin{table}[htb]
\caption{Energy shifts and radial wave functions at the origin
computed numerical assuming the potential in Eq.~(\ref{eq:V}).  The
parameter $a_0$ here is $1/(\mu\alpha_s)$, where $\mu=M_{KK}/2$ is the
reduced mass.  The last two columns show the result obtained when
neglecting the linear term in the potential.}
\label{1S2S3S}
\vskip 0.5cm
\begin{tabular}{cccccc}
\hline\hline
$M_{KK}$ \qquad & \qquad level \qquad  & \qquad $\Delta E$ (GeV)\qquad 
& \quad $a_0^{3/2}R(0)$ \qquad & \qquad $\Delta E_H$ (GeV)\qquad   & 
\quad $a_0^{3/2}R(0)_H$
\\\hline
$300$      &   1S   &  $-1.319$      &    $3.096$ & $-1.334$  & $3.079$ \\
$300$      &   2S   &  $-0.276$      &    $1.173$ & $-0.333$  & $1.089$ \\ 
$300$      &   3S   &  $-0.030$      &    $0.763$ & $-0.149$  & $0.593$ \\\hline
$500$      &   1S   &  $-2.213$      &    $3.085$ & $-2.223$  & $3.079$ \\
$500$      &   2S   &  $-0.521$      &    $1.122$ & $-0.555$  & $1.089$ \\
$500$      &   3S   &  $-0.171$      &    $0.670$ & $-0.248$  & $0.593$ \\
\hline\hline 
\end{tabular}
\end{table}      

For $A=1$~GeV~fm$^{-1}$ and $\alpha_{s}=0.1$, the radial Schr\"odinger
equation was solved numerically for the 1S, 2S, and 3S energy levels.
The results are listed in Table \ref{1S2S3S} for KK-quark masses of
300 GeV and 500 GeV, along with the predictions of a hydrogen-like
potential.  As expected, both the energy eigenvalues and $R(0)$ are
nearly hydrogen-like, justifying the use of Eq. (3.10) in the decay rate
calculations.

We see that the mass splittings, especially between the 1S and 2S
states, are substantially larger than the width of these states, and
thus will be discernible in a collider with sufficient energy
resolution.  We now turn to experimental detection of these states.
%
\section{Detection}\label{sec:detect}

The production cross section for KK-quarkonia at a linear collider can
now be discussed.  For definitiveness, we first consider the isosinglet
KK-quarks, assuming $1/R=500$ GeV and $\Lambda R = 20$.  The masses of the 
$d_R^{(1)}$ and $s_R^{(1)}$ are then $572.14$ GeV, the $b_R^{(1)}$ is
$572.16$~GeV,  and the $u_R^{(1)}$ and $c_R^{(1)}$ are $573.84$~GeV.
(The mass of the $t_R^{(1)}$ is actually a few GeV lighter, but its
decay signature, as noted in the last section, is completely different.)

\begin{figure}[htb]
\vskip -4.2cm
    \centerline{ \epsfxsize 4.5in {\epsfbox{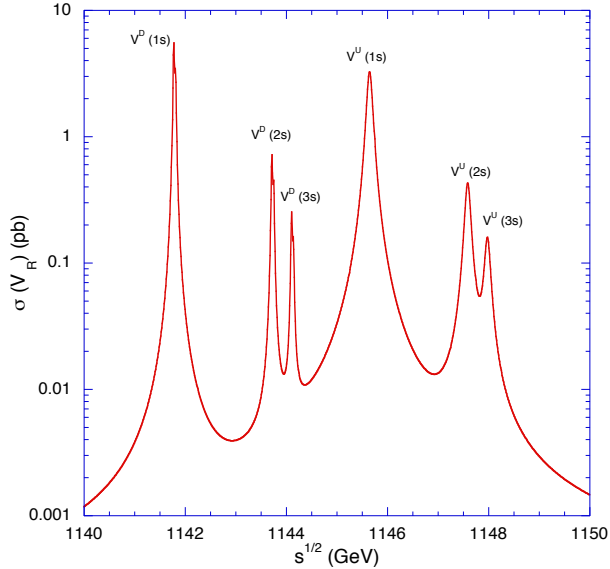}}}
 \vskip -3.6cm   
\caption{The cross section for KK-quarkonia formed by isosinglet
KK-quarks as a function of $\sqrt{s}$ for $1/R=500$~GeV and $\Lambda
R=20$.  The labels $V^D$ refer to the bound states of isosinglet
KK-down, KK-strange and KK-bottom quarks, while $V^U$ refers to the
bound states of isosinglet KK-up and KK-charm quarks.}  
\label{singletcross}
\end{figure}

Putting these together, we find the cross section as a function of
$\sqrt{s}$ given in Fig.~\ref{singletcross}.  The signature is very
dramatic--one expects two monochromatic (in this case, $67$~GeV)
quarks and large missing energy.  Clearly, the splitting between the
resonances is large enough to separate the states.  In the case of the
top KK-quark, the resulting cross section looks identical to those of
the up and charm KK-quark, but now the signature would be a
fermion-antifermion pair, each with an energy of $570$ GeV.
\begin{figure}[htb]
\vskip -4.2cm
    \centerline{ \epsfxsize 4.5in {\epsfbox{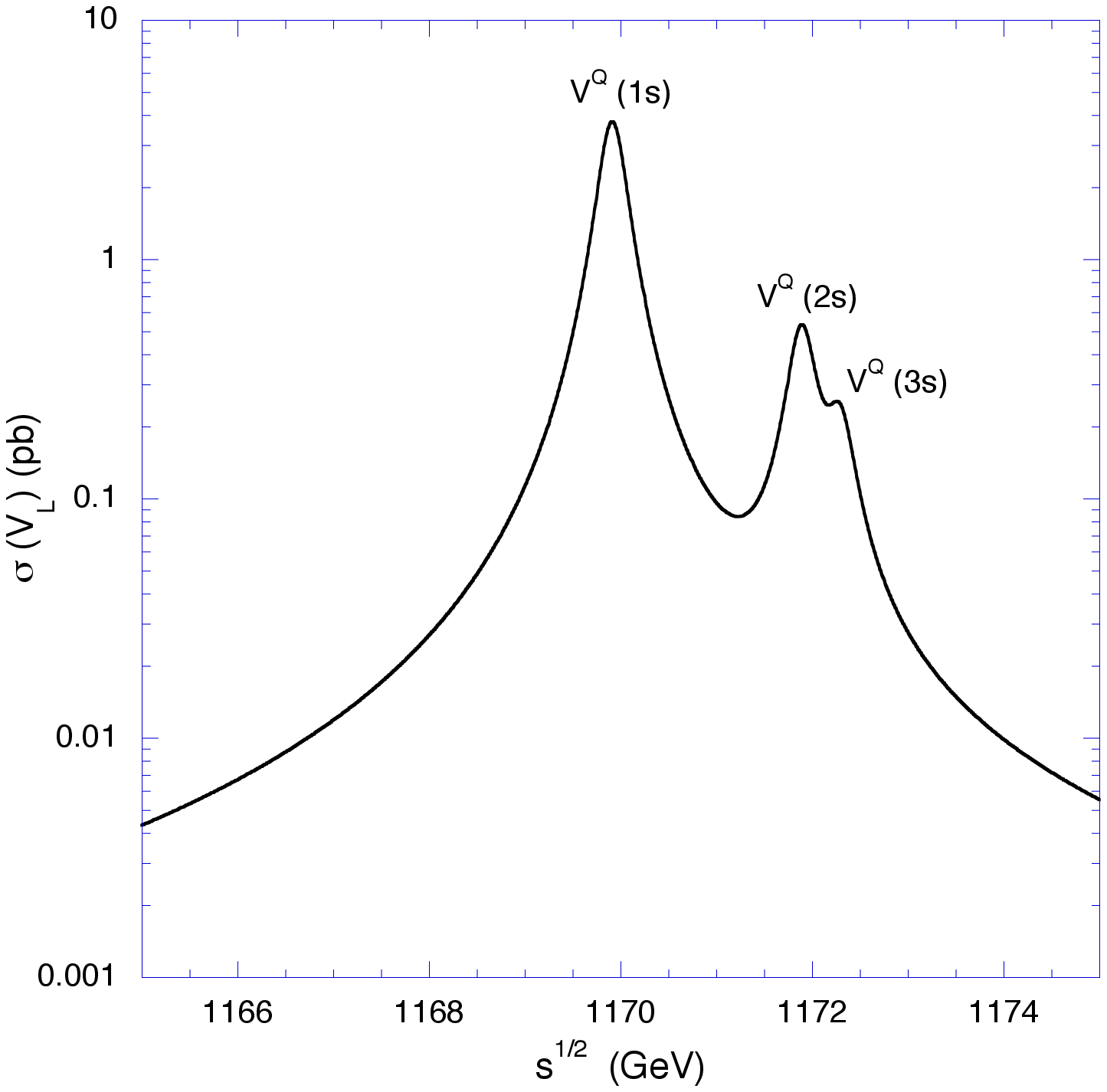}}}
 \vskip -3.7cm   
\caption{The cross section for KK-quarkonia formed by isodoublet
KK-quarks as a function of $\sqrt{s}$ for $1/R=500$~GeV and $\Lambda
R=20$. The label $V^Q$ refers to all of the isodoublet KK-quarks, 
except for the KK-top.
}  \label{doubletcross}
\end{figure} 

The masses of the isodoublet KK-quarks, except for the top, are nearly
degenerate at $585.7$~GeV.  The cross section is plotted in
Fig.~\ref{doubletcross}.  Again, the splitting between the resonances
is large enough to separate the low-lying states.  Here, the
signatures are also dramatic, with two monochromatic quarks (in this
case with energies of $46$ GeV) and, depending on the decay chain,
charged leptons, as discussed earlier.  The isodoublet top KK-quark
has a similar cross section, but is approximately 12 GeV heavier.

\begin{figure}[htb]
\vskip -5.2cm
    \centerline{ \epsfxsize 4.5in {\epsfbox{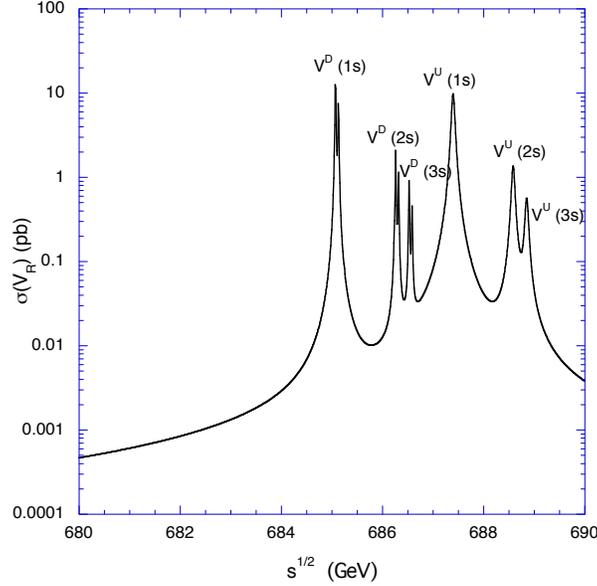}}}
 \vskip -3.7cm   
\caption{The cross section for KK-quarkonia formed by isosinglet KK-quarks 
as a function of $\sqrt{s}$ for $1/R=300$~GeV and $\Lambda R=20$.  The 
labels are the same as in the previous figures.}  \label{isosinglet300.eps}
\end{figure}

\begin{figure}[htb]
\vskip -4.0cm
    \centerline{ \epsfxsize 4.5in {\epsfbox{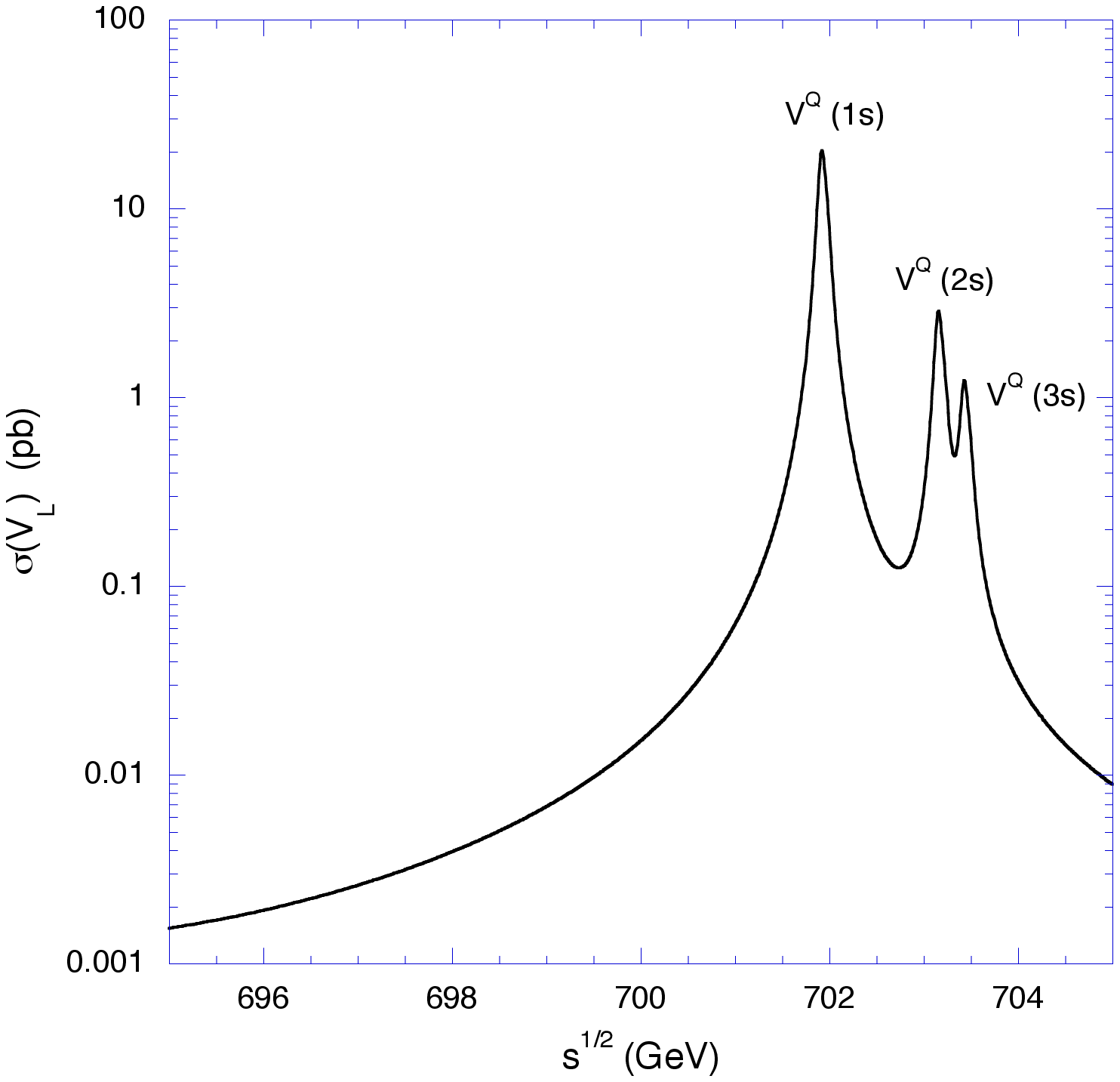}}}
 \vskip -3.7cm   \caption{The cross section for KK-quarkonia formed 
by isodoublet KK-quarks as a function of $\sqrt{s}$ for $1/R=300$~GeV 
and $\Lambda R=20$. The labels are the same as in the previous figures.}  
\label{isodoublet300.eps}
\end{figure}

These center-of-mass energies are rather high.  However, the lower
bound on the size of the extra dimensions is approximately $300$ GeV.
Using this value of $1/R$, we find the results in 
Figs~\ref{isosinglet300.eps} and \ref{isodoublet300.eps}, which are similar
to the $1/R=500$ GeV case.  Note that one can discern the fact that
the KK-bottom quark is slightly heavier than the KK-down and
KK-strange quarks, leading to some substructure in the resonances.  Of
course, in all of these cases, the $n=2$ modes will be out of reach of
a TeV scale linear collider.

Of course, it will still not be possible to detect these structures if
the beam resolution is too large.  At a muon collider, this will not
be a problem, since mass resolution of a few MeV is possible after 
deconvolution of the beamstrahlung and initial state radiation~\cite{muon}.  
Resolution is a potential problem for electron-positron colliders,
however, since one expects the average energy loss at $\sqrt{s}=500$
GeV to be approximately $1.5\%$ \cite{monig}.  This energy loss comes
from initial state radiation and beamstrahlung.  However, the
spectrum for each is well known and it is expected \cite{monig,wilson} 
that the resulting mass resolution after deconvolution will be better than
$10^{-4}$, possibly a few times $10^{-5}$, or $50$ MeV for a
$\sqrt{s}=1000$ GeV.  Such a mass resolution would easily allow the
states to be detected (although precise width measurements would
require better resolution).  Clearly, a dedicated simulation would be
needed to determine the capabilities of a linear collider (such a
simulation would also be relevant for long-lived fourth generation
quarkonia, and other s-channel resonances) for detection of
KK-quarkonia states.

\section{Conclusions}\label{sec:conc}
If the simple model of Universal Extra Dimensions that we have considered 
is realized in nature, the mass splittings between the $n=1$ KK-quarks 
and the lightest KK particle will be substantially smaller than the 
splitting between the top and bottom quarks. As a consequence, KK-quarks 
can be sufficiently long lived to form bound states, that we call 
KK-quarkonia, for a wide range of model parameters.  With boundary
mass corrections renormalized to vanish at an ${\cal O}$(TeV) cutoff scale 
$\Lambda$, we show that the KK-quark decay widths are in the 10-100 MeV 
range. We find that the peak cross sections for the 1S KK-quarkonia states 
are of the order of a few picobarns, and that the production cross sections 
near threshold show very clear and distinctive 1S, 2S and 3S resonant peaks.  
The decay signatures are very dramatic and nearly background-free: each 
isosinglet KK-quark (except the top) will decay into missing energy and a 
monochromatic quark (whose energy is determined solely by the 
KK-quark masses), and each isodoublet KK-quark will decay into missing 
energy, a monochromatic quark, and one or more leptons arising from 
subsequent two-body decays. The key issue for experimental detection is 
achieving sufficient energy resolution.  This will not be a difficulty 
for a muon collider, and not impossible for an electron-positron machine.
However, determining the resolution in the later case will require 
simulations to deconvolve the beamstrahlung and initial state radiation 
energy loss mechanisms.


\appendix
\section{Decay Width Formulae}
In this appendix, we give the partial decay width expressions for the
decays of $n=1$ KK-quarks by retaining all fermion masses and the
mixing angle $\theta_1$. For the decays of isosinglet KK-quarks, the
partial decay widths can be expressed as
\begin{eqnarray} \displaystyle
\!\!\!\Gamma(d_{R}^{(1)}\to d_{R}\,\gamma^{(1)})\!\!&=&\!\!\frac{e^2
m_{d_{R}^{(1)}}\cos^2 \theta_1}{288 \pi \cos^2 \theta_{W}}
\lambda^{1/2}(\kappa_{\gamma},\kappa_{d})\!\!
\left[\frac{1}{\kappa_{\gamma}}\lambda(\kappa_{\gamma},\kappa_{d})
+3(-\kappa_{\gamma}+\kappa_{d}+1)\right],\nonumber\\
 \!\!\!\Gamma(u_{R}^{(1)}\to u_{R}\,\gamma^{(1)})\!\!&=&\!\!\frac{e^2
m_{u_{R}^{(1)}}\cos^2 \theta_1}{72 \pi \cos^2 \theta_{W}}
\lambda^{1/2}(\kappa_{\gamma},\kappa_{u})\!\!
\left[\frac{1}{\kappa_{\gamma}}\lambda(\kappa_{\gamma},\kappa_{d})
+3(-\kappa_{\gamma}+\kappa_{d}+1)\right],\nonumber\\
 \!\!\!\Gamma(d_{R}^{(1)}\to d_{R}\,Z^{(1)})\!\!&=&\!\!\frac{e^2
m_{d_{R}^{(1)}}\sin^2 \theta_1}{288 \pi \cos^2 \theta_{W}}
\lambda^{1/2}(\kappa_{Z},\kappa_{d})\!\!
\left[\frac{1}{\kappa_{\gamma}}\lambda(\kappa_{\gamma},\kappa_{d})
+3(-\kappa_{\gamma}+\kappa_{d}+1)\right],\nonumber\\
 \!\!\!\Gamma(u_{R}^{(1)}\to u_{R}\,Z^{(1)})\!\!&=&\!\!\frac{e^2
m_{u_{R}^{(1)}}\sin^2 \theta_1}{72 \pi \cos^2 \theta_{W}}
\lambda^{1/2}(\kappa_{Z},\kappa_{u})\!\!
\left[\frac{1}{\kappa_{\gamma}}\lambda(\kappa_{\gamma},\kappa_{d})
+3(-\kappa_{\gamma}+\kappa_{d}+1)\right],
\end{eqnarray} 
and those for isodoublet case are \begin{eqnarray}
\displaystyle
\!\!\!\!\!\Gamma(d_{L}^{(1)}\!\!\to\!d_{L}\!\gamma^{(1)})\!\!\!&=&\!\!\!
\frac{e^2
m_{d_{L}^{(1)}}}{128 \pi}\!\!\!\left[\frac{1}{3}\frac{\cos
\theta_1}{\cos \theta_W}-\frac{\sin \theta_1}{\sin
\theta_{W}}\right]^2\!\!\!\!\lambda^{1/2}(\kappa_{\gamma},\kappa_{d})
\!\!\left[\frac{1}{\kappa_{\gamma}}\lambda(\kappa_{\gamma},\kappa_{d})
+3(-\kappa_{\gamma}+\kappa_{d}+1)\right],\nonumber\\
\!\!\!\!\!\Gamma(u_{L}^{(1)}\!\!\to\!u_{L}\!\gamma^{(1)})
\!\!\!&=&\!\!\!\frac{e^2
m_{u_{L}^{(1)}}}{128 \pi}\!\!\!\left[\frac{1}{3}\frac{\cos
\theta_1}{\cos \theta_W}+\frac{\sin \theta_1}{\sin
\theta_{W}}\right]^2\!\!\!\!\lambda^{1/2}(\kappa_{\gamma},\kappa_{u})
\!\!\left[\frac{1}{\kappa_{\gamma}}\lambda(\kappa_{\gamma},\kappa_{d})
+3(-\kappa_{\gamma}+\kappa_{d}+1)\right],\nonumber\\
\!\!\!\!\!\Gamma(d_{L}^{(1)}\!\!\to\!d_{L}\!Z^{(1)})\!\!\!&=&\!\!\!\frac{e^2
m_{d_{L}^{(1)}}}{128 \pi}\!\!\!\left[\frac{\cos \theta_1}{\sin
\theta_{W}}+\frac{1}{3}\frac{\sin \theta_1}{\cos
\theta_{W}}\right]^2\!\!\!\!\lambda^{1/2}(\kappa_{Z},\kappa_{d})\!
\!\left[\frac{1}{\kappa_{\gamma}}\lambda(\kappa_{\gamma},\kappa_{d})
+3(-\kappa_{\gamma}+\kappa_{d}+1)\right],\nonumber\\
\!\!\!\!\!\Gamma(u_{L}^{(1)}\!\!\to\!u_{L}\!Z^{(1)})\!\!\!&=&\!\!\!\frac{e^2
m_{u_{L}^{(1)}}}{128 \pi}\!\!\!\left[\frac{\cos \theta_1}{\sin
\theta_{W}}-\frac{1}{3}\frac{\sin \theta_1}{\cos
\theta_{W}}\right]^2\!\!\!\!\lambda^{1/2}(\kappa_{Z},\kappa_{u})
\!\!\left[\frac{1}{\kappa_{\gamma}}\lambda(\kappa_{\gamma},\kappa_{d})
+3(-\kappa_{\gamma}+\kappa_{d}+1)\right],\nonumber\\
\!\!\!\Gamma(u_{L}^{(1)}\!\!\to\!d_{L}\!W^{(1)})\!\!\!&=&\!\!\!\frac{e^2
m_{u_{L}^{(1)}}}{64 \pi\sin^2
\theta_{W}}|V_{ij}|^2\lambda^{1/2}(\kappa_{Z},\kappa_{d})
\left[\frac{1}{\kappa_{\gamma}}\lambda(\kappa_{\gamma},\kappa_{d})
+3(-\kappa_{\gamma}+\kappa_{d}+1)\right]\,,
\end{eqnarray} 
where 
\begin{eqnarray}
\lambda(x,y)=(1-x-y)^2-4xy,\,\,\kappa_{q}=\frac{M_{q}^2}{M_{KK}^2},
\,\,\kappa_{\gamma}=\frac{M_{\gamma^{(1)}}^2}{M_{KK}^2},\,\,\kappa_{Z}
=\frac{M_{Z^{(1)}}^2}{M_{KK}^2},\,\,\kappa_{W}=\frac{M_{W^{(1)}}^2}{M_{KK}^2}.
\end{eqnarray}
and $q=u,d,s,c$ or $b$.

\begin{acknowledgments}
We thank Jon Urheim, Ayres Freitas and Tao Han for useful comments.
We thank the NSF for support under Grant Nos. PHY-0140012, PHY-0243768,
PHY-0352413 and PHY-0243400. The work of IT was supported by the
Scientific and Technical Research Council of Turkey 
(T\"{U}B\.{I}TAK) 
in the framework of NATO-B1 program.  CDC thanks the William and Mary
Endowment Association for its support.
\end{acknowledgments}

       
\end{document}